\documentclass[doublecol]{epl2} 

\usepackage{hhline}
\usepackage{xcolor}

\usepackage{graphicx,subfigure,amsmath,bbm}
\usepackage{graphicx}
\usepackage{multirow}
\usepackage{hyperref}
\usepackage{placeins}
\usepackage{float}
\usepackage{marvosym}

\title{Entropic competition in polymeric systems under geometrical confinement}
\shorttitle{Entropic competition in polymeric systems under geometrical confinement} 

\author{Arash Azari \inst{1,2} \thanks{\email{arashazari@gmail.com}}
 \and Kristian K. M\"uller-Nedebock \inst{1,2} \thanks{\email{kkmn@sun.ac.za}}}
\shortauthor{Azari, M\"uller-Nedebock}

\institute{                    
  \inst{1} National Institute for Theoretical Physics, Marais Street, Stellenbosch, 7600, South Africa \\
  \inst{2} Institute of Theoretical Physics/Department of Physics, Stellenbosch University, Matieland 7602, South Africa
}


\abstract{
Using molecular dynamics simulation, we investigate the effect of confinement on a system that comprises several stiff segmented polymer chains where each chain has similar segments, but length and stiffness of the segments vary among the chains which makes the system inhomogeneous. The translational and orientational entropy loss due to the confinement plays a crucial role in organizing the chains which can be considered as an entropy-driven segregation mechanism to differentiate the components of the system. Due to the inhomogeneity, both weak and strong confinement regimes show the competition in the system and we see  segregation of chains as the confining volume is decreased. In the case of strong spherical confinement, a chain at the periphery shows higher angular mobility than other chains, despite being more radially constrained.}

\begin{document}

\maketitle

\section{Introduction }

Spatial organization is one of the key features in nature and the evolution of species where, in general, the balance between size, shape, efficiency, environmental parameters, and energy consumption does matter \cite{harold2005molecules}. In other words, one expects and does see many examples of spatial organization, crowding, and geometrical confinement everywhere, especially inside the living cell where we have large number of components within a very compact space; examples include the DNA compaction and multiple chromosomes organization inside the nucleus of the cell where  very long biopolymers are organized inside a very small space  \cite{zhou2008,cremer2001,Lanctot,Bickmore},  assembly or disassembly of proteins like polypeptide chain in chaperone \cite{hartl2002,MitBest2008}, besides some theoretical examples of polymer confinement \cite{Kindt20112001,Jun15082006,Cook21092009,TakaElie2006,Taka,Nano2006,SuAxel2007,MorThir2009,smyhar,Gao,halverson}. In addition to the theoretical and biological implications, these ideas are relevant to the nanodevices and their fabrication  \cite{wu2004composite,claessens,Xu2014,Reisner}. 

\begin{figure}[h]
\includegraphics[width=.45\textwidth]{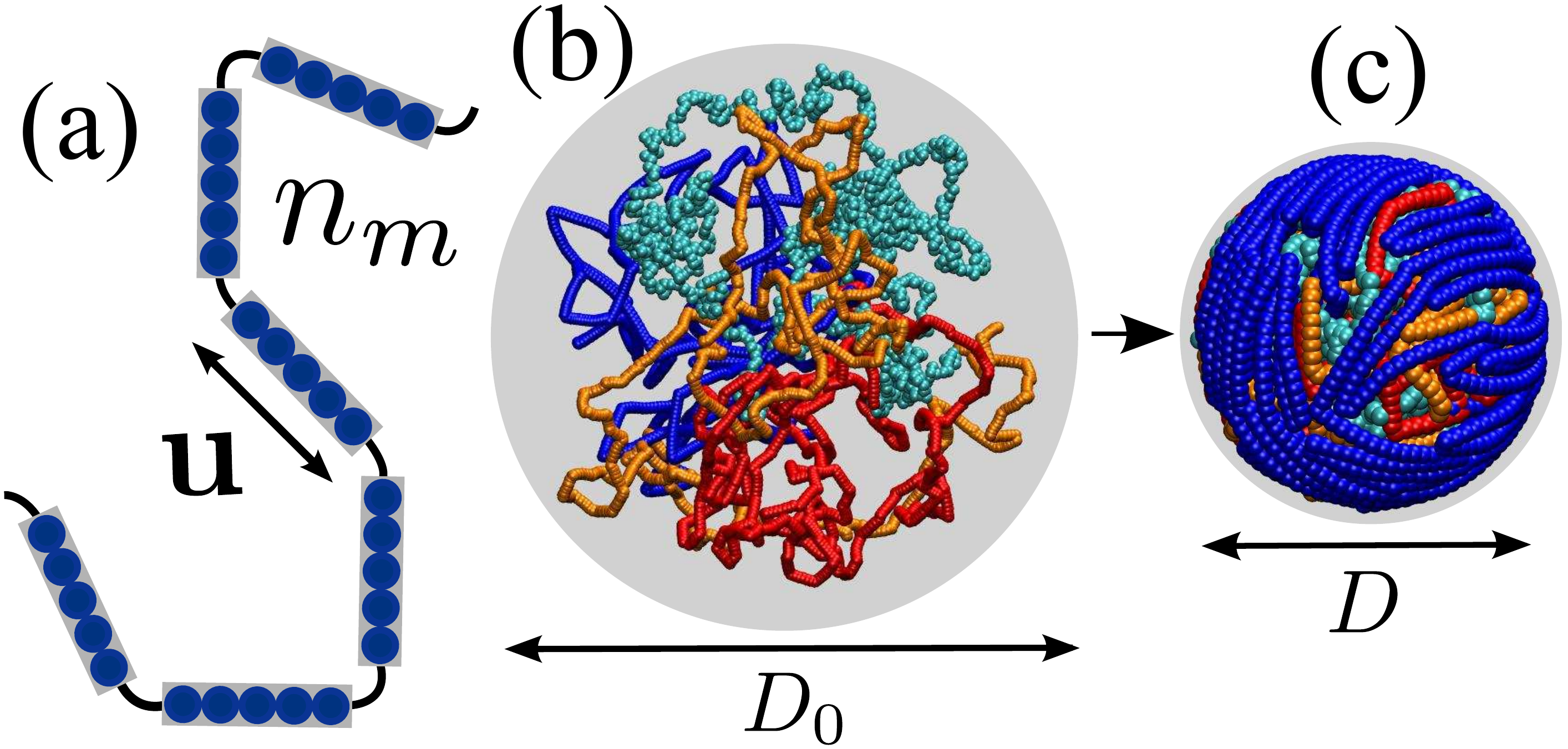}
\caption{(color online) Schematic representation of the segmented chains and the confining geometry: (a) stiff segmented polymer consists of linear close-packing of united monomers $n_{m}$, which creates the segment, and freely jointed segments with flexible bonds. The segments behave as rod-like polymer. ${\bf u}$ is the segment orientation vector. (b) initial snapshot of the simulation where $D_{0}/2$ is larger than the average radius of gyration of each chain. We have a mixture of polymers with different number of monomers per segment.(c) the final snapshot; here $D/2$ is smaller than the radius of gyration of each chain. The figure schematically shows  strong confinement: longer segments (in blue) are pushed to the periphery.} \label{fig:seg_poly}
\end{figure}

The confinement of multiple homopolymer chains shows different behavior associated with dilute to overlapping regime, equivalently from weak to strong confinement   \cite{Grosberg94a,TakaElie2006}. In overall, as density is increased for system of homogeneous and identical chains, these chains lose their identity and all the physical and thermodynamical quantities are equally likely for each chain; e.g. monomer density, chain size, and so on. By introducing inhomogeneity, specially in the form of multiple chains, where different chains have different structural characteristic, similar to the scenario inside the nucleus where we have many different components with different physical or chemical properties, addressing the phase behavior and thermodynamical properties of the system becomes very complicated; examples include crowding inside the nucleus, chromosome condensation and formation of chromosome territories \cite{cremer2001,Lanctot,bancaud2009,crowd}.

It has been presented that geometrical confinement can induce phase separation \cite{gelbrev,Jun15082006}. Despite its simplicity, Flory-Huggins theory (FH) \cite{Huggins41,Flory42} needs to be improved to be able to deal with phase behavior of more complex systems of polymer blend where we have monomers with structure, stiffness, and different molecular weights \cite{phaseJCP,ACS,freed2005,PRISM}. FH theory simply indicates that the mixing is not favorable and in general, a  polymer blend is immiscible. 

Here, we would like to address the competitive entropic segregation in a mixture of different polymer chains under geometrical constraint using molecular dynamics simulation. Importantly the chains differ in how they are segmented, rather than interactions or total numbers of monomers. We introduce the inhomogeneity to the system by using different stiff segmented chains, Fig. \ref{fig:seg_poly}(a), and keeping all the interactions the same between all chains. Segment represents a group of united monomers which behaves as rod-like polymer, Fig. \ref{fig:seg_poly}(a). There are some similarities between our polymer chains and some coarse-grained models of main-chain liquid crystal elastomers \cite{de_gennes}, euchromatin and heterochromatin structures inside the nucleus \cite{Gaspar}, and lengthwise condensation mechanism \cite{marko}, besides some theoretical models \cite{Flory1978,Kristian2003}. 

\section{Simulation and the model}

The simulations, in the $(N,V,T)$ ensemble, were performed using the ${\rm ESPResSo}$ package \cite{Limbach2006704}. The length scale in our simulation is the diameter of each monomer, $\sigma$, and we scale the energies based on the Lennard-Jones (LJ) energy $\epsilon$. We consider the Lennard-Jones time scales as a measurement unit for the time $\tau=\sigma \sqrt{m/\epsilon}$, where $m$ is set to unity for all the monomers. We used the Langevin thermostat with damping constant $\gamma=\frac{1}{2}\tau^{-1}$ to keep the equilibrium temperature of the system around $T=\epsilon/k_{\mathrm B}$. The velocity Verlet algorithm was used with time step of $t=0.001\tau$ during the equilibrating stage, and $t=0.01\tau$ for production run. Here we considered $\sigma=1$, $\epsilon =1 (k_{\mathrm B}T)$, and temperature is $T=1$.

We construct the polymer chains as bead-spring model which is confined in a large spherical container with impenetrable wall, Fig. \ref{fig:seg_poly}(b). For each chain, we use the prefix {\em intra} for anything related to the united monomers in each segment, and {\em inter} where it indicates between the segments of the same chain.  To mimic the excluded-volume effect, a purely repulsive Lennard-Jones potential is applied to the bead-bead and bead-wall (container) interactions, $V_{\rm LJ}\left(r\right)=4\epsilon \left[ \left( \sigma/r\right)^{12} -\left( \sigma/r\right)^{6} +1/4 \right] $ if $r < 2^{\frac{1}{6}} \sigma$, otherwise it is zero. The potential has been truncated and shifted. The bonded interactions between the beads, both inter and intra interactions, are finitely extensible nonlinear elastic (FENE) potential to keep the equilibrium distance between the beads (monomers) around $r_{0}= \sigma$, $V \left( r \right)= -\frac{1}{2} K_{\mathrm F} \Delta r_{\mathrm {max}}^{2} \ln \left[ 1-\left( \left(r-r_{0}\right)/\Delta r_{\mathrm {max}} \right)^{2} \right]$, where $\Delta r_{\mathrm {max}}$ is the maximal bond stretching and $K_{\mathrm F}$ is the spring constant. Besides that, to make rod-like stiff segments, we used the cosine squared bond angle potential between united monomers in each segment which increases the intra-stiffness, the bending potential, $  V \left( \phi \right)= K_{\mathrm a}/2 \left[\cos\left(\phi\right)-\cos\left(\phi_{0}\right) \right]^{2} $, $K_{\mathrm a}$ is the bond angle bending constant and $\phi_{0}=\pi$ is the equilibrium bond angle.

The total number of monomers per each polymer chain is $N_{m}=1365$ where we have $4$ chains in each simulation. Each chain is composed of serial connection of similar segments (united monomers). By changing the number of monomers per segment $n_{m}$, Fig. \ref{fig:seg_poly}(a), and changing the intra-stiffness (persistence length) of the segments, we create flexible, normal-stiff short, extra-stiff short, and extra-stiff long segments where the flexible, short, and long  segments have $n_{m}=1$, $7$, and $15$ monomers per segment respectively. The flexible chain is a normal self-avoiding chain with FENE bond between each monomer. 

For the chains with normal stiff segments, we used the following value for the FENE bond spring constant $K_{\mathrm F}=200 \:\frac{k_{\mathrm B} T}{\sigma^{2}}$, and the bending constant for bond angle potential is $K_{\mathrm a}=200 \:\frac{k_{\mathrm B} T}{\sigma^{2}}$. For the chains with extra stiff segments, we used $K_{\mathrm F}=400 \:\frac{k_{\mathrm B} T}{\sigma^{2}}$ and $K_{\mathrm a}=600 \:\frac{k_{\mathrm B} T}{\sigma^{2}}$ for FENE and bond angle potentials respectively. The flexible bond between segments, inter-stiffness, is a FENE bond with $K_{\mathrm F}=15 \:\frac{k_{\mathrm B} T}{\sigma^{2}}$. The maximal bond stretching was set to $\Delta r_{\mathrm {max}}=1.5\:\sigma$ for all FENE bonds.

The normal-stiff segment has persistence length of $l_{\rm p}\approx15\sigma$, where it is $l_{\rm p}\approx27\sigma$ for  extra-stiff segment. The persistence lengths of the chains made by these normal-stiff short and extra-stiff long segments are $l_{\rm p}\approx3.8\sigma$ and $7\sigma$ respectively. The average length of the long and short segments are $11.6 \sigma$ and $5.2 \sigma$ respectively. The average unconfined radii of gyration, $\langle R_{\rm g}\rangle $, of flexible, short, and longer segment chains are $33\sigma$, $36\sigma$, and $42 \sigma$ respectively. We define the orientation vector ${\bf u}$ as the vector which connects the first monomer to the last monomer in each segment. For flexible chain it simply represents the bond between successive monomers. We started with a large spherical container with radius larger than the average radius of gyration of each polymer chain, $R=49\sigma$ Fig. \ref{fig:seg_poly}(b), which was followed by contraction to $30\sigma$, $20\sigma$, and $14\sigma$ radii, Fig. \ref{fig:seg_poly}(c). Therefore, the  volume fractions for different confinement radii are $\varphi=4 N_{m} \sigma^{3}/ D^{3}= 0.0058, 0.025, 0.085, 0.25$, respectively.

In the simulations of each radii of confinement we performed equilibration, production run plus sampling, then shrinking the volume in a very slow process which lets the system relax the tensions and prevents the chains to strongly entangled to each other and becomes trapped into one of the local energy minima of the system.  In overall we performed 24 separate simulations. 

We equilibrate the system for $2\times 10^{6}\,\tau$ time steps and the production run was $3.7\times 10^{9}\,\tau$ time steps (the whole production run). We sampled the configurations each $200\,\tau$ time steps. Shrinking the volume happens during a very slow process, $1.5\times 10^{6}\,\tau$ steps. In case of strong confinement, $R=14\sigma$, we run extended simulations for additional $3.7\times 10^{9}\,\tau$ time steps to make sure the chains are relaxed and the results are not artifact of the simulation under strong confinement.

\begin{figure}[h]
\includegraphics[width=.48\textwidth]{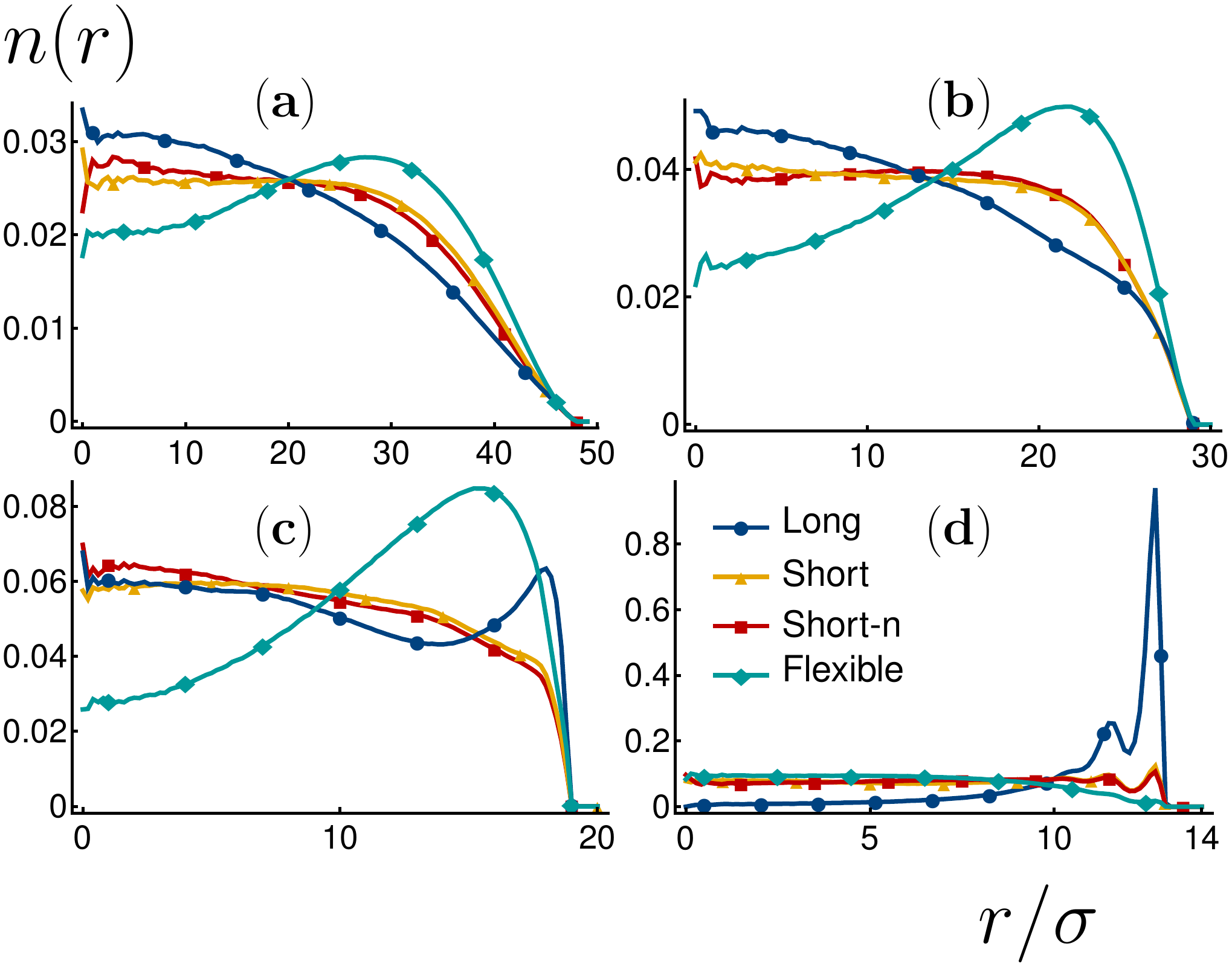}
\caption{(color online) The number density distribution of the monomers inside the sphere with different degrees of confinement. The radius of the confining sphere in each case is (a) $49$, (b) $30$, (c) $20$, and (d) $14$ all in unit of $\sigma$. Here we have a mixture of extra-stiff long segments (Long), extra-stiff short (Short), normal-stiff short (Short-n), and Flexible polymer chains. By stiff, we mean intra-stiffness. Entropic effect shows how segregation occurs in both weak and strong confinements. For weak confinement, the flexible chain moves toward the boundary, (a-c), and for strong confinement, (d), we see segregation of longer segment chain at the periphery. The markers do not represent the whole data points, solely used for differentiation of curves. } 
\label{fig:N_dens}
\end{figure}

\section{Results} The monomer number density, $n\left(r\right)$, is defined based on the location of each monomer relative to the center of the sphere. We divide the volume into the concentric shells with thickness of $\Delta r=R/100$ and average the number of monomers in each shell. Fig. \ref{fig:N_dens} shows the normalized number densities for different confinement radii. In weak confinement where $R=49>\langle R_{\rm g}\rangle$, Fig. \ref{fig:N_dens}(a), positioning of the segments near the boundary is not favorable as it increases the non-bonded interaction which is accompanied by higher entropic loss of peripheral conformation for longer segments; longer segments lose more translational and orientational entropies close to the boundary than other chains. Therefore we can see in figures \ref{fig:N_dens}(a,b) that longer segments avoid the periphery, instead, the flexible chain with less entropic penalty cost moves towards the boundary.

In contrast, the more strongly confined state in Fig. \ref{fig:N_dens}(c) drives the system to the point that the entropic competition forces the longer segments to start relocating to the periphery where they have less bending energy cost. Further confinement, Fig. \ref{fig:N_dens}(d), shows segregation of longer segment chain; all the longer segments congregate at the periphery where they adopt the nematic orientation. It is disentangled from the rest of the chains and forms a clump of segments where they can move together with less bending energy cost and relatively lower {\em translational} entropy loss. This characteristic entropy-driven segregation and differentiation is something that we could call {\em entropic chromatography} \cite{ent_chro}.  A short video of the simulation trajectory is also included as Supplementary Video.

Experiments on actin filaments\cite{claessens}, mixture of actin and DNA \cite{Negishi} which are confined in small emulsion droplet, also confirm the segregation state and formation of a cortex close to the surface of the droplet under strong confinement; similar to the segregated chain at the periphery in our simulations, Fig. \ref{fig:N_dens}(d). 

The number density plots, Fig. \ref{fig:N_dens}, show when the persistence length of the segment is greater than or comparable to the length of the segment, extra intra-stiffness has a minor effect as short-normal and short-stiff segments behave similarly. By increasing the degree of confinement, Fig. \ref{fig:N_dens}(a-c), the flexible chain is pushed more towards the boundary due to the lowest entropic cost and interestingly, except for the strong confinement case, in all other cases its number density close to the center remains relatively unchanged. For strong confinement we also see that the number densities for shorter segments and flexible chains are similar except for the behavior close to the wall where the flexible chain is strongly suppressed, but there is still a finite probability for shorter segments to be at the periphery.  

In order to make sure that the segregation is not artifact of the simulation and the polymer chains are not trapped due to the decrease in the radius of confinement, from $R=20\sigma$ to $14\sigma$, in strong confinement case, $R=14\sigma$, we removed all the bond angle potentials (set the bond angle bending constant to $K_{\mathrm a}=0$) to make all the chains as flexible polymer chains with the same FENE potential; making a system of homopolymers. We let the chains to reach the equilibrium state where the number densities of monomers of each chain show similar pattern which indicates there is no competition between the chain. Then we started to increase the $K_{\mathrm a}$ during a very slow process of $4\times 10^{6}\,\tau$ time steps. We continued the simulations for $1\times 10^{9}\,\tau$ time steps and the results are the same as before without any changes which confirms the segregation is not an artifact of the simulations.

We performed separate simulations on systems of identical homopolymers, only similar stiff-segmented chains, and there is no segregation in those systems and we have a uniform number density distribution regardless of the length of the segment. This implies the importance of inhomogeneity in competitive phase segregation \cite{Axel}. 

In another series of simulations, we connect all the chains together to create a very long single chain in all possible combinations. Regardless of the chain order, we find the same result as for not connected chains. This indicates that connectivity plays a minor role and always the longer segments are pushed to the periphery.

\begin{figure}[h]
\includegraphics[width=.48\textwidth]{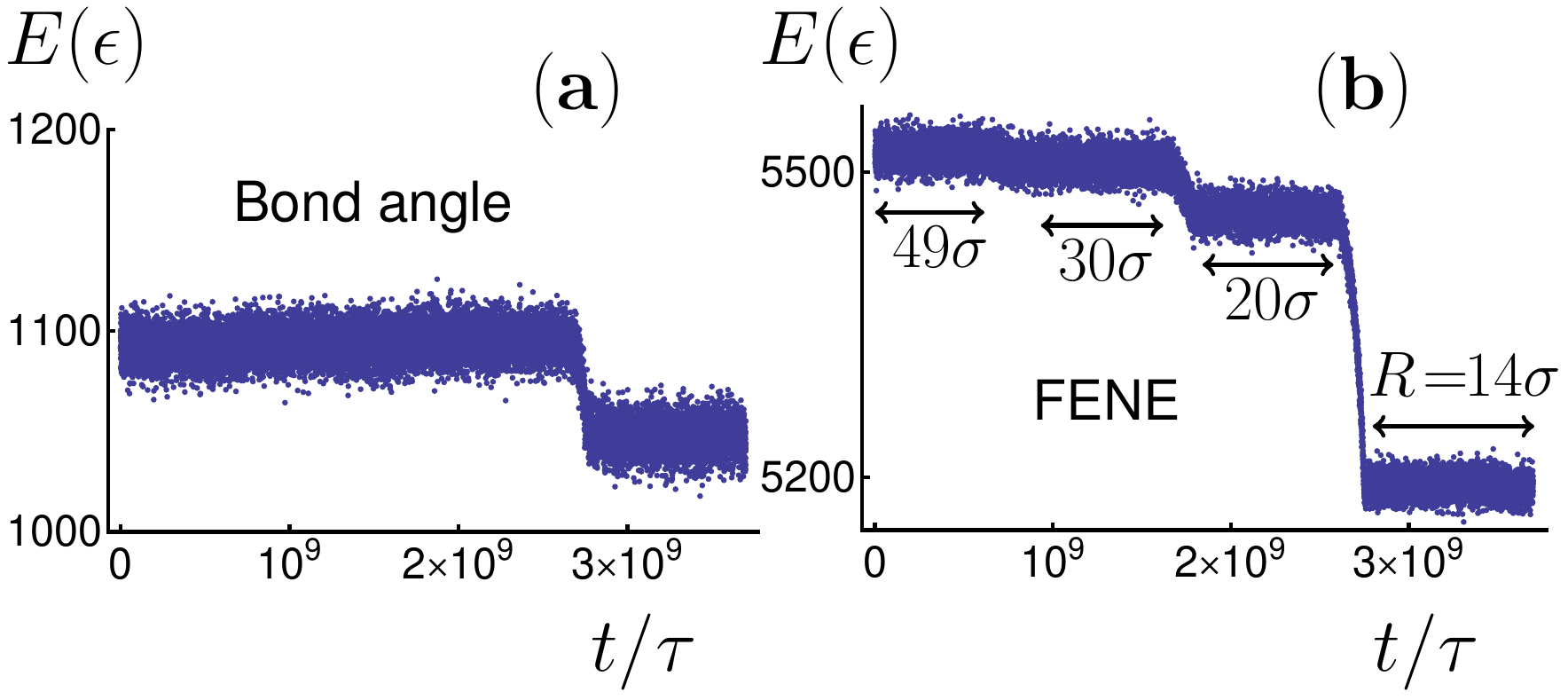}
\caption{(color online) The potential energy contributions of bonded interactions. (a) bond angle and (b) FENE potentials. The bond angle energy plot represents the extra-stiff segment energy changes and the FENE energy plot shows the changes in the energy of the flexible bond between the segments. The plots show the stepwise increasing degree of confinement. Reorganization and segregation of the chains decrease the bonded potential energy contribution to the total energy. The radii of the confining sphere ($R$) for each plateau are represented in (b).} 
\label{fig:energy_}
\end{figure}

The energy contribution of each potential term, Fig. \ref{fig:energy_}, shows that by increasing the degree of confinement, each of the bonded interaction contributions (FENE and bond angle) is decreased and all the non-bonded energy contributions (LJ) are increased (not presented). The increase of non-bonded potential is clear as by increasing the degree of confinement, on average the particles become closer (  the system behaves like molecules of gas with high density so the non-bonded energy loosely depends on the conformation of the chains and only fluctuates around some average value). Increasing the degree of confinement results in a higher bending energy cost, but by reorganizing the chains, the system tries to minimize this penalty and disentangle the chains. Besides that, the FENE potential prevents overstretched bonds, despite the higher non-bonded repulsive force, to lower the bonded energy. As a result of cooperation between FENE,bond angle potentials, and entropic effect, the segregated phase has the lower energy state (in terms of bonded potential energy contribution). In figure \ref{fig:energy_}, we just presented two energy terms, bond angle potential of extra-stiff segments and the FENE potential of flexible bond between the segments, but the behavior of all the other terms are similar and the energy contribution of all bonded potentials are decreased by increasing the degree of confinement (see Supplemental Material (SM)\cite{SM_} for notes on the pressure in strong confinement regime).

\begin{figure}[ht]
\includegraphics[width=.48\textwidth]{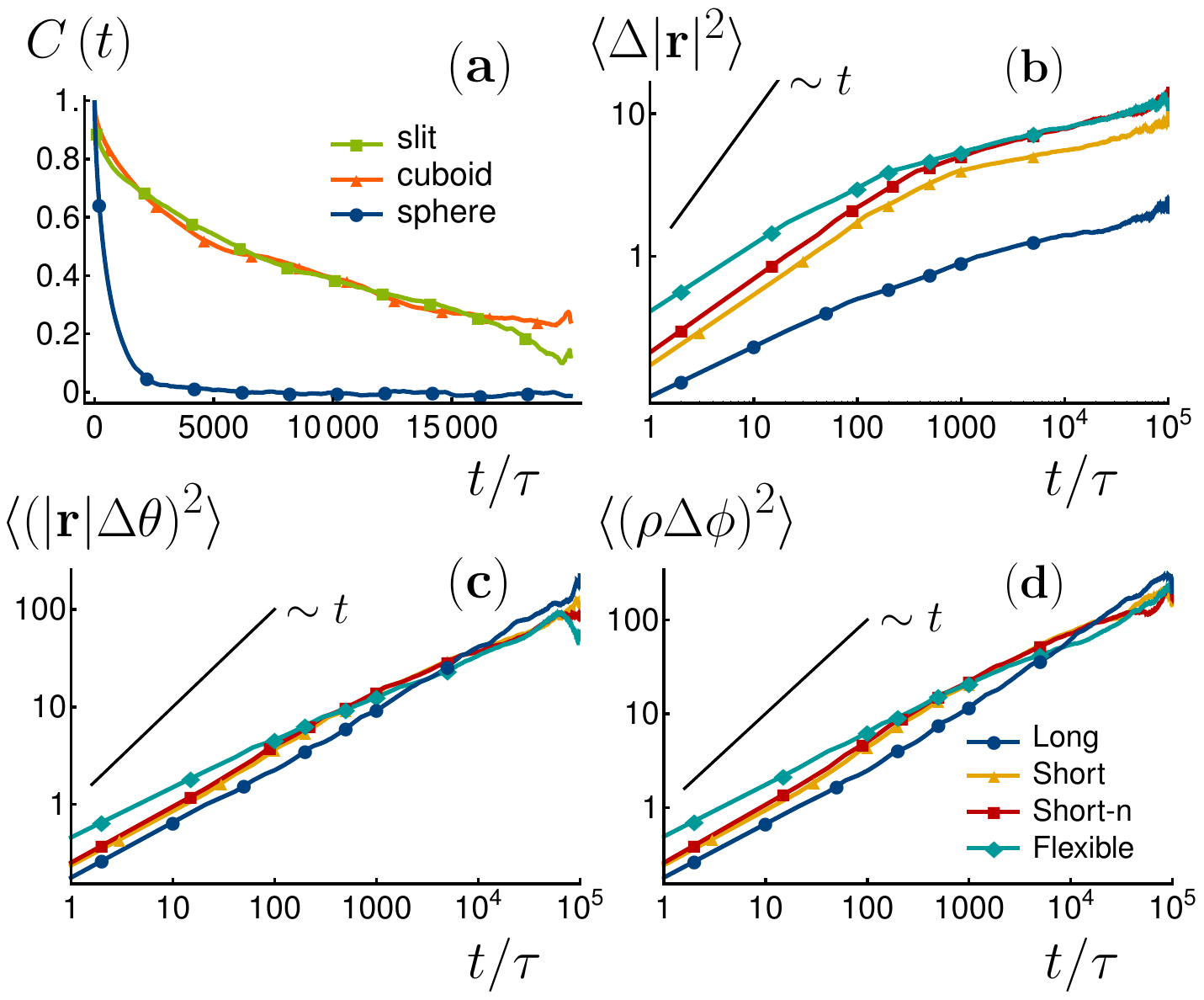}
\caption{(color online) (a) The segment orientation correlation and (b) the radial, (c) polar, and (d) azimuthal mean square displacements; all calculated under strong confinement regime. (a) represents the segment orientation correlation for longer segment chain in spherical, cuboid, and slit geometries where it shows that the segments of the chain at periphery of spherical geometry are in constant reorientation compared to other geometries. (b-d) show the MSDs for all the chains in strong spherical confinement. The angular MSDs, polar and azimuthal, show relatively higher angular mobility ($\sim 30-40\%$) for longer segment chain at periphery.} 
\label{fig:corr_msd}
\end{figure}

We investigated the effect of curvature of the container by performing simulations with the same chains and interactions in both cuboid container with impenetrable walls and slits with periodic boundary conditions \cite{aakkmn}. For both cuboid and slit we see a similar segregation of the chains at the same volume fraction as in spherical case, $\varphi= 0.25$, where the longer segments congregate near the walls \cite{aakkmn}. The main difference is, the longer segment chain at the periphery of confining sphere is more mobile than the longer segment chains at periphery of cuboid or slit geometries. Indeed, we examine this by calculating the segment orientation correlation function, $C\left(t \right)=\langle{\bf u}_{i}(t_{0}) \cdot{\bf u}_{i}(t_{0}+t)\rangle_{i,t_{0}}$, for longer segments in strong confinement regime. The result indicates the longer segments at periphery of the sphere are in a constant reorientation and relocation, but in other geometries they tend to maintain their orientation, Fig. \ref{fig:corr_msd}(a).

In the strong confinement regime, we calculate the mean square displacement (MSD) of every monomer in each chain in long-time limit \cite{dowards,diff_conf}, $\langle \Delta A(t)^{2}\rangle=\langle ( A_{i}(t_{0}+t)-A_{i}(t_{0}))^{2}\rangle_{i,t_{0}}$. We calculate radial $\langle \Delta |{\bf r}|^{2} \rangle$, azimuthal $\langle \left(\rho \Delta \phi\right)^{2} \rangle $, and polar $\langle \left(|{\bf r}| \Delta \theta \right)^{2} \rangle $, MSDs where $\rho=|{\bf r}| \sin (\theta)$. All of them show behavior similar to the constrained diffusion, Fig. \ref{fig:corr_msd}(b-d). The longer segment chain at the periphery, despite the suppression of its radial displacement, represents higher angular (both azimuthal and polar) mobility compared to the other chains ($\sim 30-40\%$). The long-time limit slopes of all the MSDs are represented in table \ref{tab.1} . The higher angular mobility is in agreement with the interpretation that the segregated chain is a clumped structure which is disentangled from the rest of the chains so it has relatively higher angular mobility (diffusivity) at periphery \cite{explain2}.

\begin{table}
\caption{Long-time limit slopes of the mean square displacements (MSD) for each chain in the system. The long-time slopes of the radial $r$, polar $\theta$, and azimuthal $\phi$, MSDs are reported for each chain in strong confinement regime.}
\label{tab.1}
\begin{center}
\begin{tabular}{|c||c|c|c|c|}
\hline
MSDs & Long  & Short & Short-n & Flexible \\
\hhline{|=||=|=|=|=|}
$r$ & $0.16$ & $0.15$ & $0.19$ & $0.17$ \\
\hline
$\theta$ & $0.64$ & $0.42$ & $0.39$ & $0.42$ \\
\hline
$\phi$ & $0.73$ & $0.51$& $0.48$& $0.42$ \\
\hline
\end{tabular}
\end{center}
\end{table} 

\begin{figure}[ht]
\includegraphics[width=.48\textwidth]{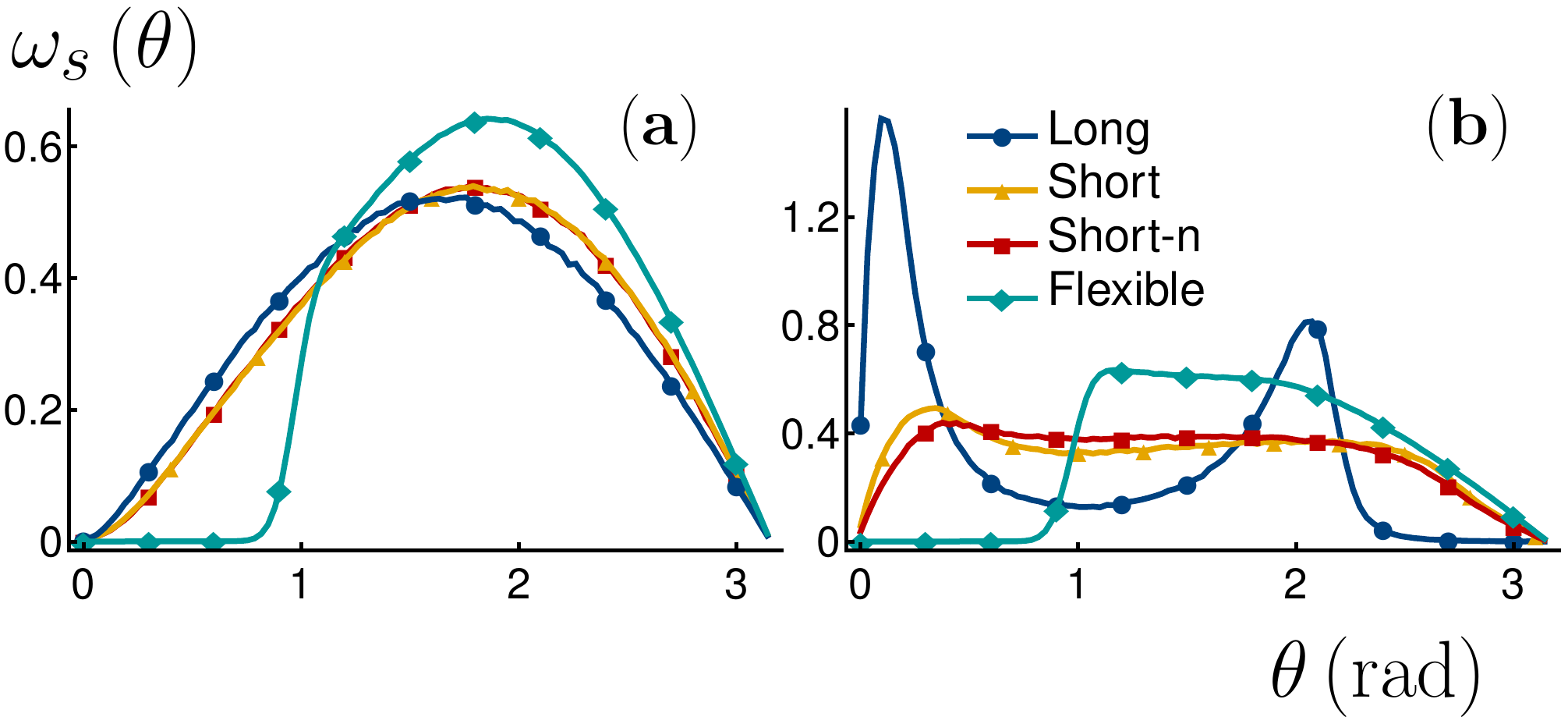}
\caption{(color online) The segment-segment angle frequency distribution. The plots represent distribution of the angle between two successive segments for each chain. The radius of the confining sphere in each plot is (a) $49 \, \sigma$ and (b) $14 \, \sigma$ which shows a drastic change in the angle distribution of longer segments where they are mainly parallel to each other, the first peak, or forming part of an imaginary polygon.} 
\label{fig:bond-ang-dist}
\end{figure}

The segment-segment angle distribution gives us information about the orientation of the segments relative to each other which can be calculated based on the angle between two successive segments $\theta =  \langle \arccos \left( {\bf u}_{i}(t)\cdot {\bf u}_{i+1}(t)/\left|{\bf u}_{i}(t) \right| \left|{\bf u}_{i+1}(t) \right| \right)\rangle_{i,t} $. The results are presented in Fig. \ref{fig:bond-ang-dist}.

In weak confinement, Fig. \ref{fig:bond-ang-dist}(a), where the balance between excluded volume interaction and entropy of each chain makes it swell, we see a distribution of angles which by increasing the degree of confinement, tends slightly towards the smaller angles for longer segments, see SM \cite{SM_}. The distribution for flexible chain is indeed the bond angle distribution and due to the purely repulsive interaction between every other monomer, they cannot approach each other closer than a permitted length $\sigma$ so the angle distribution for flexible chain has a threshold around $\pi/3$. Strong confinement: Fig. \ref{fig:bond-ang-dist}(b) shows how longer segments create nematic orientation as we see only two main angle distribution peaks. The initial peak indicates relatively parallel alignment of a majority of consecutively joined segments and the second peak resembles formation of mainly hexagon or nonagon by longer segments, see SM \cite{SM_}. For shorter segments we see kind of homogeneous distribution with slightly higher possibility of finding the segments in folded state. The flexible chain does not show any significant changes in different confinement regimes. Experiments \cite{Lettinga} also confirm the effect of geometrical confinement on the reorientation of colloidal rods.

\section{Conclusion}

We proposed a simple coarse-grained stiff segmented polymer model and we applied only the excluded volume effect as non-bonded interaction between the monomers. Our results represent an entropic competition which depends on the degree of confinement where under strong confinement, we see entropy-driven segregation of the chains. The entropy behaves as a tool in any system which represents structural inhomogeneity to organize itself without consuming external energy. The entropy-driven segregation potentially may find some applications in systems where separation of polymer blend is desirable and manipulation is difficult or limited.  Instead, it would be highly advantageous to assign the separation to the system to reorganize itself. We would refer to bio/polymers under confinement \cite{claessens}, nanodevice fabrication plus microfluidics and micro-encapsulation \cite{DeCock,Watanabe,Wang,Seiffert} besides implications for chromosomes organization inside the nucleus of the cell \cite{Lanctot,Jun15082006,Cook21092009}.

The nucleus of the eukaryotic cell is a good example of a crowded inhomogeneous system under geometrical confinement where besides many proteins and enzymes, there are two major states of chromatin; euchromatin and heterochromatin. Euchromatin represents a more flexible chain of nucleosomes which is an active gene-rich chromatin and heterochromatin is a dense stiff compaction of nucleosomes and considered as inactive gene-poor chromatin \cite{Gaspar}. Depending on the cell type, the diameter of the nucleus is around $2-10 \mu {\rm m}$, therefore, the volume fraction of the chromatin inside the nucleus varies between $2\%-30\%$ \cite{halverson,Phillips}. If we compare our stiff segmented polymer model, Fig. \ref{fig:seg_poly}(a), and the coarse-grained structures of euchromatin and heterochromatin \cite{Gaspar}, we can draw a qualitative analogy between them where our segments represent the heterochromatin and the rest of the flexible parts represent the euchromatin. The volume fraction in our system changes from $0.5\%$, weak confinement, to $25\%$ for strong confinement regime; comparable to the nucleus volume fraction. It has been confirmed that the heterochromatin in normal eukaryotic cells is mainly positioned at the  periphery and anchored to the nuclear lamina by tethering and euchromatin is mainly at interior space ready for gene expression and replication \cite{Lanctot,Bickmore,Akhtar}. Our results, qualitatively represent a mechanism which can organize and position the segments based on their structural properties, i.e. stiffness and length. In case of strong confinement regime, this entropy-driven segregation mechanism pushes the longer stiff segments, heterochromatin, towards the periphery where they can find a proper tethering site, and more flexible chains, euchromatin, are located at interior space; making distinction between components of the system. Our findings qualitatively show that besides all other active metabolism for nucleus organization \cite{Lanctot,Bickmore}, the entropic effect potentially can play an important role in nucleus organization \cite{Jun15082006,Cook21092009,Axel}.

Higher mobility of the chain at the periphery and mobility of other chains show another possible implication for biological systems like chromosomes inside the nucleus where during their recombination process, gene expression, or search for tethering site on nuclear lamina, the components should diffuse not merely by active metabolism as the curvature and entropic effect can contribute and facilitate these processes \cite{Hubner,Marshall,Judrod}.

\acknowledgments
We would like to thank anonymous referees for their fruitful comments and Axel Arnold, Minne P. Lettinga, Davide Marenduzzo, and Marco Polin for very useful discussions. This research was conducted using the computational resources of Centre for High Performance Computing (CHPC) and Stellenbosch University, Rhasatsha HPC.

\section{\huge Supplemental Material}

\section{Field theoretical analogy}

To elaborate the origin of the entropic competition and phase behavior, we present a brief field theoretical description \cite{Glenn} of the main-chain liquid crystalline polymers (LCPs) \cite{de_gennes} which are similar to our stiff segmented polymers, Fig. 1(a) main article.  We consider the LCPs as wormlike chains, although this model assumes the whole chain as a semiflexible chain which resists to the bending, it provides simple explanation for LCPs behavior \cite{Glenn}. The microscopic density of segment orientation and position of $n$ wormlike polymer chains is defined as $\hat{\rho} \left({\bf r},{\bf u} \right)\equiv\sum_{j=1}^{n}\int_{0}^{L_{C}}\mathrm{d}s \; \delta \left({ \bf r}-  {\bf r}_{j}\left(s \right) \right) \delta \left({ \bf u}-  {\bf u}_{j}\left(s \right) \right)$. The arc length of the polymer is defined by $s\in \left[0,L_{C} \right]$ where $L_{C}$ is the contour length of the polymer. The configuration of the chain is represented by ${\bf r}_{j}\left(s \right)$, therefore, ${\bf u}_{j}\left(s \right) =\mathrm{d}{\bf r}_{j}/\mathrm{d}s$ is the tangent vector to the polymer at $s$. The canonical partition function of a system consists of main-chain nematic wormlike polymers is ${\cal Z}_{C}\left(n,V,T \right)={\cal Z}_{0} \int {\cal D}\rho \int {\cal D} w\; \exp \left(-H\left[ \rho,w\right] \right)$, where ${\cal Z}_{0}$ is the partition function for an ideal gas of $n$ non-interacting wormlike chains which can be regarded as configurational entropy and $H$ is the effective Hamiltonian 

\begin{equation}
\begin{split}
 H\left[ \rho,w\right]&=-i \int \mathrm{d}{\bf r} \int \mathrm{d}{\bf u}\; w \left({\bf r},{\bf u} \right) \rho \left({\bf r},{\bf u} \right) -n \ln Q\left[ i w\right] \\ 
 & \hspace{-1cm} +\frac{\beta}{2} \int \mathrm{d}{\bf r} \int \mathrm{d}{\bf u}\int \mathrm{d}{\bf u^{\prime}}\; \rho \left({\bf r},{\bf u} \right) v \left({\bf u},{\bf u^{\prime}} \right) \rho \left({\bf r},{\bf u^{\prime}} \right)\label{eq:Hamiltonian}
 \end{split}
\end{equation}

The first term in this Hamiltonian is the interaction of each monomer with the complex chemical potential field $ i w \left({\bf r},{\bf u} \right)$. We can simply interpret $w \left({\bf r},{\bf u} \right)$ as the response of test polymer chain to the potential of all the interactions involved in the system. The second term, $Q\left[ i w\right]$ is the normalized partition function of a wormlike chain which is the entropic term \cite{Glenn}  
\begin{equation}
 Q\left[ i w\right]=\frac{\int{\mathcal{D}{\bf r}}\;\exp \left(-\beta U_{0}\left[{\bf u}  \right] -\beta U_{1}\left[{\bf r},i w  \right] \right)  }{ \int{\mathcal{D}{\bf r}}\;\exp \left(-\beta U_{0}\left[{\bf u}  \right]  \right) }
\end{equation}

\noindent where $U_{0}$ is the bonded potential between each monomer and $U_{1}$ is the interaction of each monomer with the complex chemical potential field $i w \left({\bf r},{\bf u} \right)$.

Due to the nature of the liquid crystals, the non-bonded interaction between the segments, the third term in Eq.(\ref{eq:Hamiltonian}), should depend on the orientation of the segments, $v \left({\bf u},{\bf u^{\prime}} \right) \propto \lvert {\bf u}\times{\bf u}^{\prime} \rvert$, which is the Onsager model for LCPs \cite{Onsager}. The entropic term depends on the bonded potential for each chain, by increasing the stiffness, the translational entropy loss increases which implies that in the limiting cases of flexible and rod polymers, we have minimum and maximum translational entropy loss respectively. We should note that there are complicated competitive terms in the Hamiltonian between non-bonded interaction and the entropic term which are captured by free energy minimization. The translational entropy term, $Q\left[ i w\right]$ is the only term which represents the connectivity of the polymers, therefore the spatial non-locality only arises from this term which indicates in confined geometry, we have to apply the boundary conditions to the propagator of a single-chain and geometrical boundary condition has nothing to do with other terms in the effective Hamiltonian \cite{Glenn}. We can conclude that in confined geometries, the entropic term is the dominant factor which tries to minimize the free energy of the system.

\section{Multiple similar chains; no segregation}

\begin{figure}[H]
\begin{center}
\includegraphics[width=.48\textwidth]{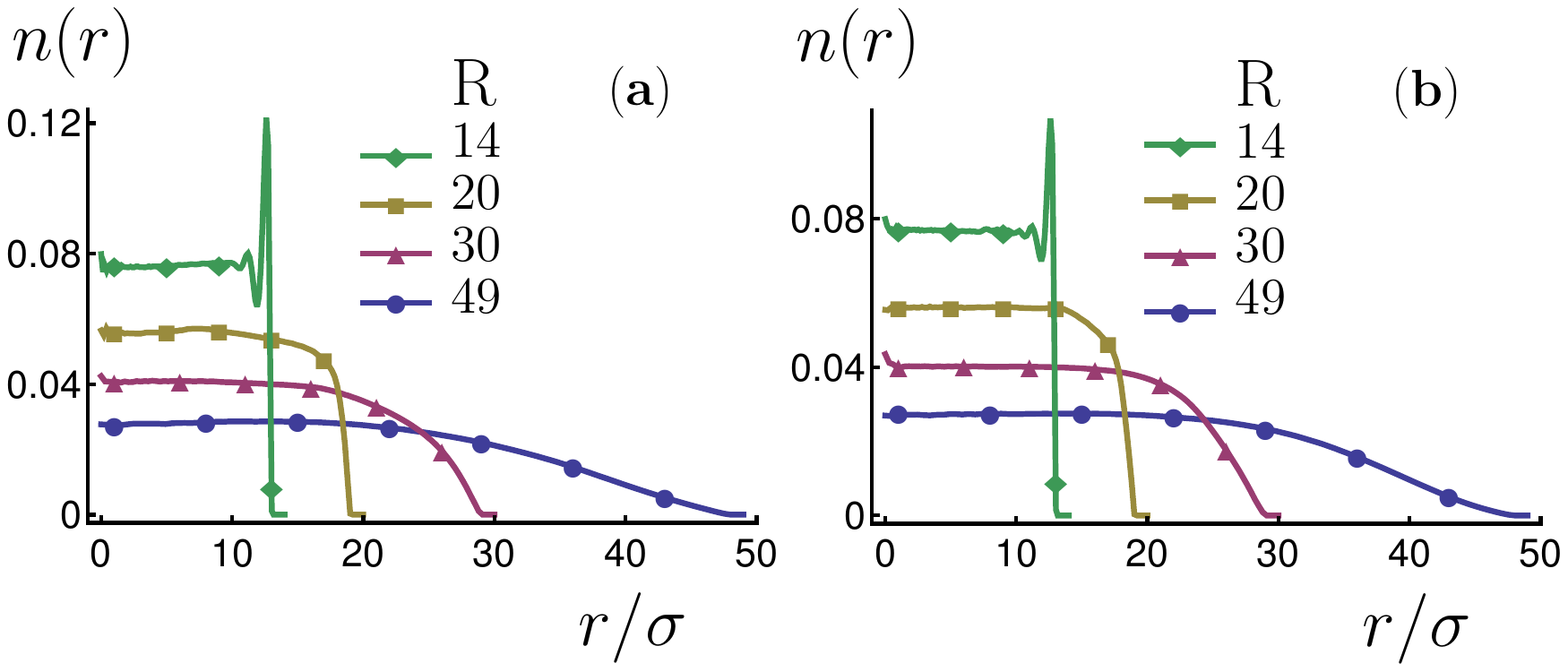}
\caption{(color online) monomer number density for homogeneous systems. (a) long segment polymers (b) short segment polymers. For homogeneous system where all the chains are homopolymers, there is no competition between the chains and both short and long segment chains behave similarly. The fluctuation (oscillation) close to the boundary resembles the solvation force and liquid layering adjacent to the solid surface where we can distinguish the contact and midpoint densities. The surface forces the monomers to reorient themselves which creates the fluctuation \cite{Israelachvili}.} \label{fig:homogeneous}
\end{center}
\end{figure}

In separate series of simulations, we used systems of homopolymers which has only 4 similar chains (long segment, or short segment, or flexible chain) and as we can see the monomer number density distribution is homogeneous through the volume and there is no segregation of the chains, Fig. \ref{fig:homogeneous}, except some fluctuation close to the boundary which represents that the chains are in a constant reorientation \cite{Israelachvili}.

\section{segment-segment angle frequency distribution }

\begin{figure}[ht]
\begin{center}
\includegraphics[width=.48\textwidth]{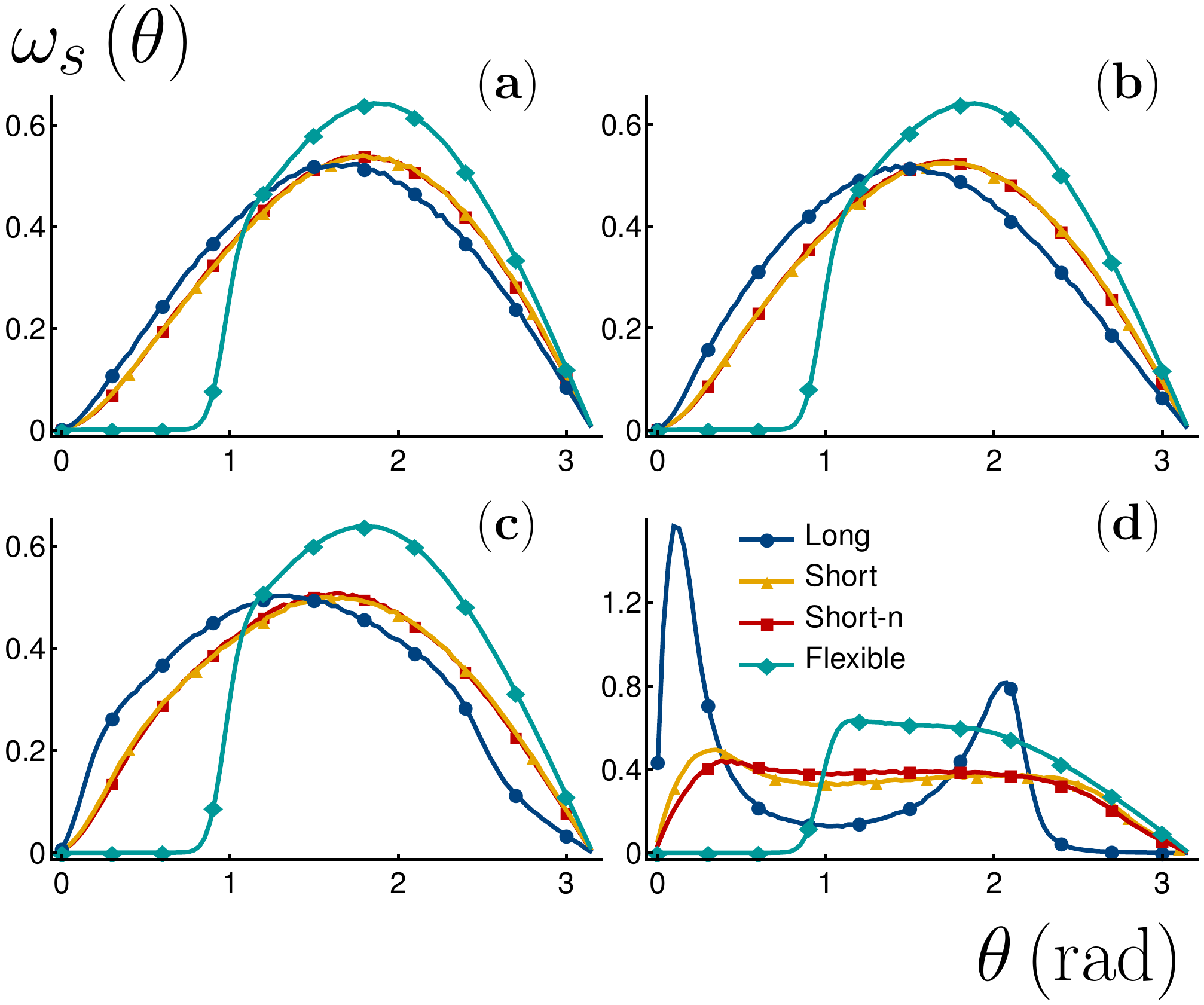}
\caption{(color online) The segment-segment angle frequency distribution. The plots represent the angle between two successive segments for each chain. The radius of the confining sphere in each case is (a) $49$, (b) $30$, (c) $20$, and (d) $14$ all in unit of $\sigma$. Figures (a-c) show the tendency of the chains to avoid very small or straight angles; we see a distribution which does not change significantly for different radii of confining geometry. In contrast, figure (d) shows a drastic change in the angle distribution of longer segments where in their nematic phase, the majority of them are parallel to each other, the first peak, or forming sides of an imaginary polygon. The rest of the chains show similar behavior in all confinement regimes.} 
\label{fig:bond-ang-distSM}
\end{center}
\end{figure}

Figure \ref{fig:bond-ang-distSM}, represents the segment-segment angle frequency distribution for all confining regimes. Figures \ref{fig:bond-ang-distSM} (a-c) do not show any significant change in the angle distribution. In contrast, the strong confinement shows a drastic change in the distribution of the angle between the longer segments, Fig. \ref{fig:bond-ang-distSM}(d). The rest of the chains show similar behavior as weak confinement regime. The distribution for flexible chain is indeed the bond angle distribution and due to the purely repulsive interaction between every other monomer, they cannot approach each other closer than permitted length $\sigma$ (center to center distance which creates a equilateral triangle) beyond which they are repelled strongly. Therefore the angle distribution for flexible chain has a threshold around $\pi/3$. Figure \ref{fig:bond-ang-distSM} also provides information about the alignment of the segments; the chains avoid very small or straight angles specially in weak confinement. The only exception is the behaviour of the longer segments under strong confinement where they can adopt parallel alignment, Fig. \ref{fig:bond-ang-distSM}(d).  

\section{segment orientation relative to radial unit vector }

The orientation of the segments is calculated based on the cosine angle between the radial unit vector, ${\bf \hat {r}}=\frac{\bf r}{|\bf r|}$, and the vector which connects the first monomer to the last monomer in each segment ${\bf u}$, with the average length $\langle \lvert{\bf u}\rvert \rangle \sim 11.6$ and $\sim 5.2\, \sigma$ for long and short segments, respectively. The result is the direction cosine angle $\cos\left(\theta \right)=\langle\left| {\bf u}_{i}(t) \cdot {\hat{\bf r}}/|{\bf u}_{i}(t)| \right|\rangle_{i,t}$. 

\begin{figure}[ht]
\begin{center}
\includegraphics[width=.48\textwidth]{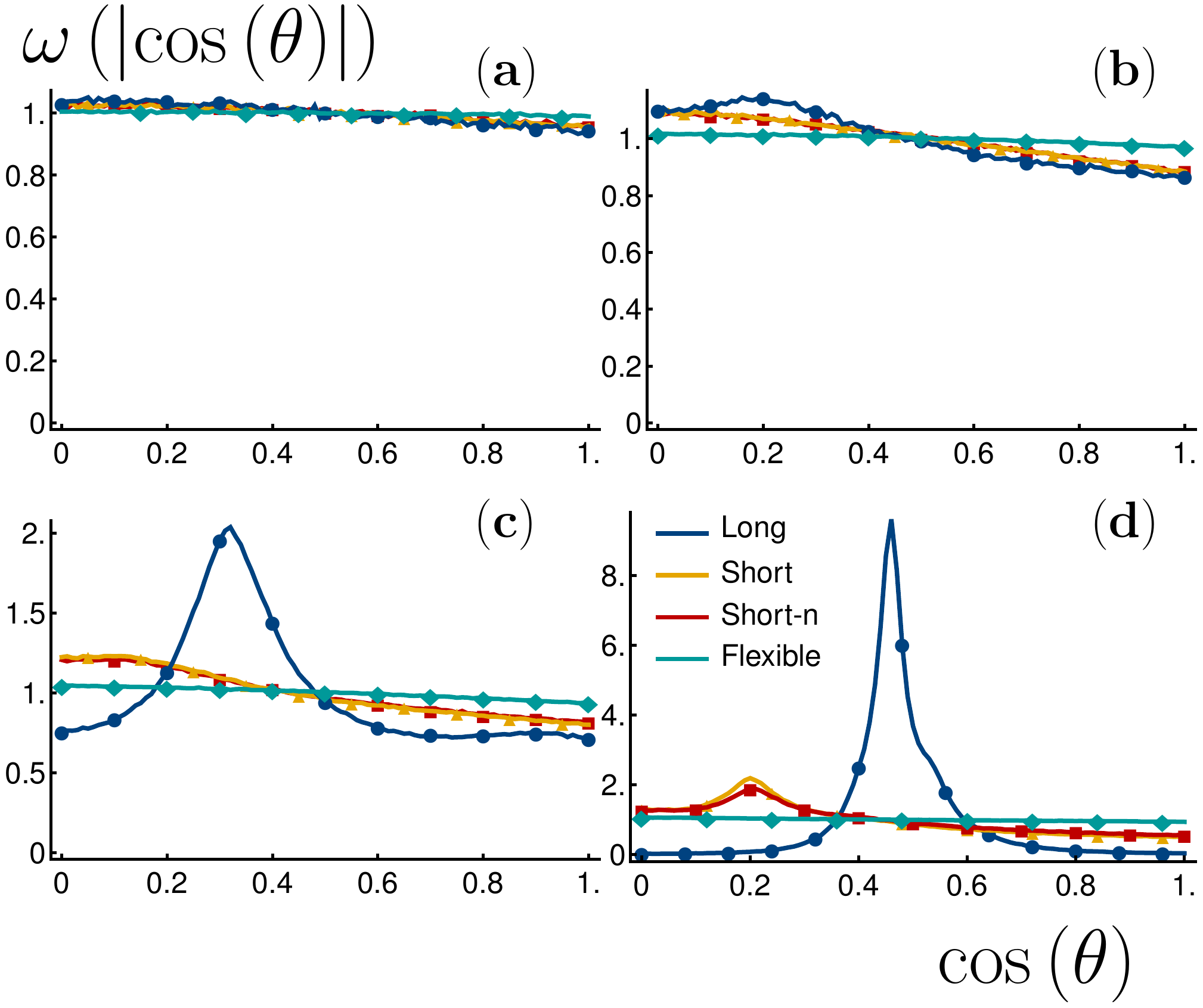}
\caption{(color online) The angle distribution between each segment and the unit radial vector. The radii of confinement are (a) $49$, (b) $30$, (d) $20$, and (e) $14\; \sigma$. For weak confinement (a-b), all the angles are equally likely, but for strong confinement (c-d), most of the longer segments take the same conformation which resembles sides of a hexagon which is circumscribed by the confining sphere. Shorter segments are less affected by confinement and the effect is trivial for flexible chain.} \label{fig:cos_angle}
\end{center}
\end{figure}

\begin{figure}[H]
\begin{center}
\includegraphics[width=.4\textwidth]{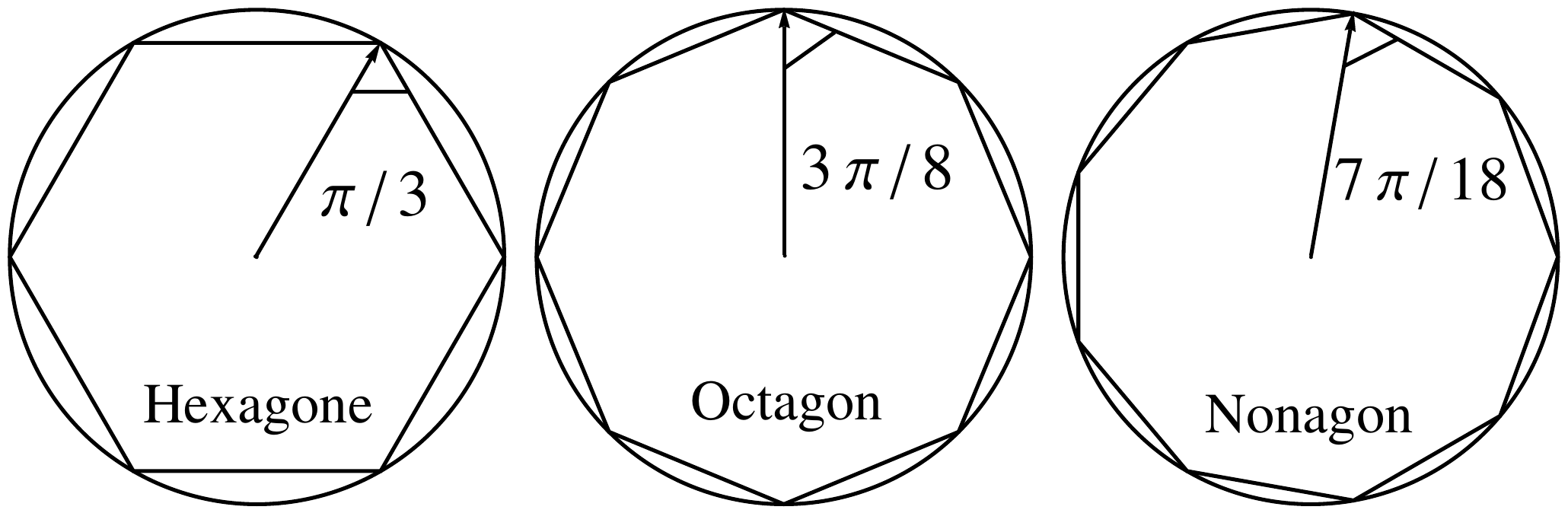}
\caption{The angle between each side of the regular polygon and the radial vector. These data are in good agreement with our findings in figure \ref{fig:cos_angle}. We did not include the octadecagon here.} \label{fig:polygons}
\end{center}
\end{figure}

For weak confinement in figure \ref{fig:cos_angle}(a-b), nearly all the angles are equally likely. Figure \ref{fig:cos_angle}(c) shows that orientation of each long segment is close to the nonagon structure (necessarily, it does not imply the formation of unique polygon or nonagon in the system, each segment get the posture as side of a polygon which is circumscribed by the confining sphere) where the angle between each side, segment, is $\frac{7\pi}{9}$ Fig. \ref{fig:polygons}. By increasing the degree of confinement, we see in figure \ref{fig:cos_angle}(d) the angles for long segments are mainly distributed around $\frac{\pi}{3}$, a shift from nonagon to hexagon Fig. \ref{fig:polygons} which again represents formation of nematic ordered conformation. For shorter segments in strong confinement regime, again we can see a polygon formation, octadecagon type, but only some of the segments are contributed in the polygon posture and occurrence of other angles is probable. For flexible chain nearly all the angles are equally distributed in all confinement regimes.

We see a good agreement between the polygon sides and the circumscribed circle of radius $\displaystyle R=\frac{s}{2\sin \left(\frac{180}{n} \right)}$, where $s$ is the length of the each side of the regular polygon. Here $s$ is the average length of the orientation vector $\langle \lvert{\bf u}\rvert \rangle$, and n is the number of sides Fig. \ref{fig:polygons}.

\section{pressure under strong confinement regime}

We measured the pressure during the simulations; in terms of reduced units (mentioned in the manuscript) the unit of pressure in our simulation is proportional to $[P]=[{\epsilon}/{\sigma^3}]=[{1k_{B} T}/{\sigma^3}]$

where $\epsilon$ is the Lennard-Jones energy ($= 1k_{B} T$) and $\sigma$ is the diameter of each monomer. 
The average pressure that we measured for strong confinement regime in the simulations is around ~ 0.75 (reduced unit).
Normally, in the coarse-grained simulations $\sigma$ is around ~$ 0.5 -1 {\rm nm}$. (We can even consider the thickness of dsDNA for diameter of the monomers ($\sigma$) which is around $2 {\rm nm}$; it does not affect the order of magnitude of the pressure in our simulations.) If we multiply the average measured pressure (0.75) by the Boltzmann constant times the temperature and divide them by the cube of $\sigma$ ($=1 {\rm nm}$) we get:$P=3.1065\times 10^{6} {\rm Pa} \sim~ 3 {\rm MPa}$
which has the same order of magnitude of the reported experimental measurements on viral capsid which is around 6 Mpa \cite{DNA1,DNA2}.


\FloatBarrier

\end{document}